\documentclass[%
 twocolumn,
 amsmath,amssymb,
 aps,
 prb,
 showpacs,
 preprintnumbers,
]{revtex4-2}

\usepackage[english]{babel}
\usepackage{graphicx}
\usepackage{dcolumn}
\usepackage{bm}
\usepackage{amsmath}
\usepackage{amssymb}
\usepackage{physics}
\usepackage{siunitx}
\usepackage{hyperref}
\usepackage{cleveref}
\usepackage{amsfonts} 

\begin{document}

\title{Wide-angle high-performance photodetector empowered \\by angle-insensitive Tamm plasmon polariton}

\author{Yurii V. Konov$^1$}
\author{Rashid G. Bikbaev$^{1,2}$}
\author{Feng Wu$^3$}
\author{Ivan V. Timofeev$^{1,2}$}  

\affiliation{$^1$Kirensky Institute of Physics, Federal Research Center KSC SB RAS, 660036 Krasnoyarsk, Russia}
\affiliation{$^2$Siberian Federal University, 660041 Krasnoyarsk, Russia}
\affiliation{$^3$School of Physics and Optoelectronic Engineering, Guangdong Polytechnic Normal University,  510665 Guangzhou, China}


\date{\today}

\begin{abstract}
Tamm plasmon-polaritons (TPPs) — optical modes localized at the interface between a metal and a photonic crystal (PhC) — offer a versatile platform for confining light in planar optoelectronic devices. However, their implementation in angle-sensitive applications such as photodetectors and solar cells is hindered by strong angular dispersion of light. In this work, we propose a strategy to overcome this limitation by tailoring the dispersive properties of a PhC through the integration of hyperbolic metamaterials (HMMs). Using the transfer matrix method and effective medium theory, we demonstrate that the HMM exhibits type-I hyperbolic dispersion in the telecommunication wavelength range. This enables a photonic bandgap whose angular dependence compensates for the intrinsic blue shift of the TPP mode, effectively anchoring the resonance at 1550 nm over a broad range of incidence angles. Device performance is evaluated using the Fowler internal photoemission model, yielding a normal-incidence responsivity of 17.5 mA/W. Notably, for TM-polarized light, the responsivity decreases by only 10\% at a 60$^\circ$ incidence angle — a substantial improvement over conventional all-dielectric PhC structures, which exhibit a degradation exceeding 86\%. Our findings establish HMM-engineered TPPs as a promising platform for wide-angle high-performance photodetectors and open new directions for dispersion engineering in active plasmonic and optoelectronic devices.
\end{abstract}

\maketitle

\section{Introduction}

The ability to confine and manipulate light at subwavelength scales is fundamental to modern photonics and optoelectronics. Among the various strategies for achieving optical confinement, surface electromagnetic modes have attracted particular attention due to their ability to concentrate electromagnetic energy at interfaces. Surface plasmon polaritons (SPPs) ~\cite{barnes2003surface, zayats2005nano, pitarke2007theory}], which propagate at metal-dielectric interfaces, have been extensively studied and applied in sensing ~\cite{doi:10.1021/acsami.0c12525}, waveguiding~\cite{10.1063/1.2008385}, and nonlinear optics~\cite{10.1063/5.0061726}. 
However, SPPs suffer from inherent Ohmic losses and require specific phase-matching conditions for excitation, typically via prism coupling or grating structures.

An alternative class of surface modes, Tamm plasmon polaritons (TPPs), was theoretically predicted by Kaliteevski \textit{et al.} in 2007~\cite{Kaliteevski2007} and subsequently experimentally demonstrated~\cite{Sasin2008Tamm}. 
TPPs are optical states localized at the interface between a metal and a one-dimensional photonic crystal (PhC). Unlike SPPs, TPPs can be excited directly by both TE- and TM-polarized light at normal incidence without additional coupling elements.

The physical origin of TPPs can be traced to the pioneering work of I. E. Tamm on electronic surface states in solid-state physics~\cite{Tamm1932}. 
Tamm showed that localized electronic states can exist at the boundary of a periodic potential with wave functions that decay exponentially into both the crystal and adjacent medium. 
The optical analogue emerges from the formal equivalence between the Helmholtz equation for electromagnetic waves in layered media and Schrödinger’s equation for electrons in periodic potentials. 
In the optical case, metal provides a "potential barrier" with a negative dielectric permittivity. 
The PhC provides a periodic potential, which results in a localized state at the interface between them. 
The unique properties of TPPs have enabled diverse applications across photonics. 
The unique properties of TPPs have enabled diverse applications across photonics. TPP was used in lasers~\cite{Symonds2013,Xu2021}, photodetectors~\cite{Konov2024, Li2014, Huang2023, Yen2025, Konov2025}, beam steering ~\cite{ma15176014,photonics10101151,Bikbaev2024} and solar cells~\cite{ Zhang2013,Bikbaev2022}.

Despite these advances, a fundamental limitation of TPP-based devices remains their inherent angular sensitivity. 
As the incidence angle increases, the PhC band gap shifts towards shorter wavelengths (blue shift), dragging the TPP resonance with it. 
This angular dispersion severely limits the operational angular acceptability of TPP-based devices. 
This is particularly problematic in applications such as photodection, imaging and solar energy harvesting, where light comes from a wide variety of angles. 
Overcoming this limitation requires engineering the dispersive properties of PhCs to compensate for angular dependence.

Hyperbolic metamaterials (HMMs) have emerged as a powerful platform for controlling light-matter interactions through their unique dispersion characteristics~\cite{Poddubny2013Hyperbolic, Ferrari2015Hyperbolic}. 
HMMs are strongly anisotropic media with dielectric permittivity tensor components of opposite signs along orthogonal directions ($\varepsilon_\parallel \varepsilon_\perp < 0$). 
The integration of HMMs into PhCs has been explored in several recent studies. 
Xue \textit{et al.}~\cite{Xue2016} demonstrated that a PhC containing HMM layers can exhibit an angle-insensitive band gap for TM polarization, where the spectral position remains constant over a wide angular range. 
Wu \textit{et al.}~\cite{Wu2018, Wu2023} extended this concept, deriving analytical conditions for achieving red shift band gaps that shift towards longer wavelengths with increasing angle for TM polarization. 
These properties arise from competition between ordinary angular dispersion in dielectric layers and extraordinary dispersion in HMM layers.

In this work, we propose HMM-based dispersion engineering to realize an angle-insensitive TPP and demonstrate its application in a wide-angle high-performance photodetector operating at the telecommunication wavelength of 1550 nm.

\section{Theoretical model}
\label{sec:structure}

\subsection{Geometry of the Proposed Structure}

\begin{figure*}[t]
    \centering
    \includegraphics[width=\linewidth]{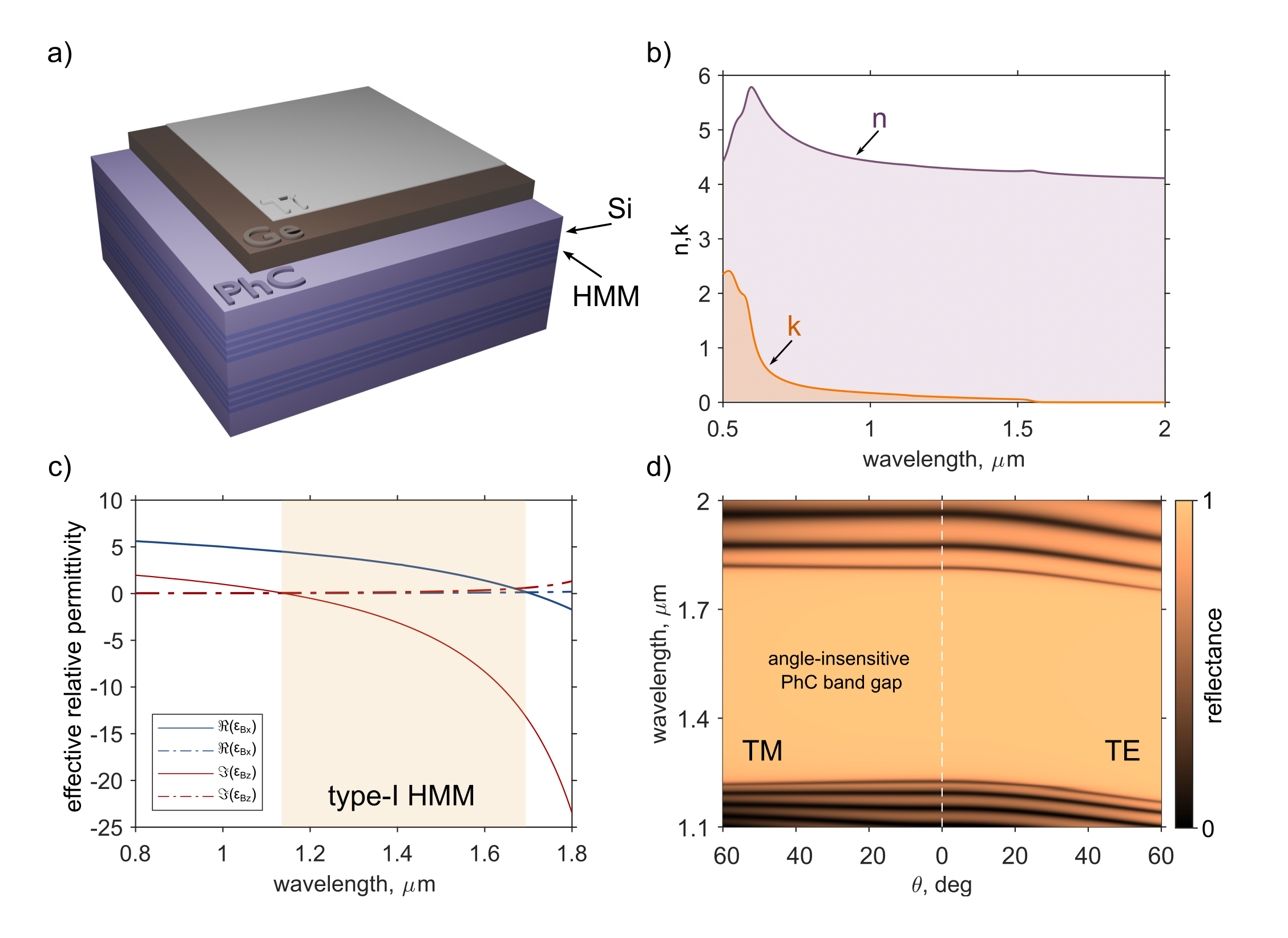}
    \caption{(a) Schematic of the proposed angle-insensitive-TPP-based photodetector. The structure consists of a Ti film, a Ge absorbing layer, and a one-dimensional PhC with alternating Si and HMM layers. The HMM comprises four periods of ITO/Si bilayers. (b) Wavelength dependence of the real refractive index $n$ and extinction coefficient $k$ of Ge, reproduced from Ref.~\cite{Nunley2016}. (c) Real parts of the effective permittivity components $\varepsilon_\mathrm{Bx}$ and $\varepsilon_\mathrm{Bz}$ of the ITO/Si HMM calculated using effective medium theory [Eqs.~\eqref{eq:eps_parallel} and \eqref{eq:eps_perpendicular}]. The purple-shaded region indicates the type-I hyperbolic regime [$\Re(\varepsilon_\mathrm{Bx})>0$, $\Re(\varepsilon_\mathrm{Bz})<0$]. (d) Reflectance spectra of the HMM-based PhC  as a function of wavelength and incidence angle for TM and TE polarizations are calculated for proposed structure presented in subplot (a).}
    \label{fig:fig_HMM_1}
\end{figure*}

Figure~\ref{fig:fig_HMM_1}(a) presents a schematic view of the proposed hot-electron photodetector. 
The structure consists of three functional blocks: (i) a top titanium (Ti) layer for exciting TPP and generating hot electrons, (ii) a germanium (Ge) \cite{Nunley2016} layer for carrier collection, and (iii) a HMM-based PhC for angular dispersion engineering.

The Ti~\cite{Johnson1974} layer thickness is $d_{\mathrm{Ti}} = 7$~nm. 
Titanium was selected due to its favorable Schottky barrier height with germanium and its good adhesion properties. 
This thickness of ultrathin film is comparable to or smaller than the mean free path of electrons in Ti ($\sim$10–15 nm), ensures that a significant fraction of photogenerated hot electrons can reach the Ti/Ge interface prior to thermalization.
Beneath the Ti layer lies a Ge film with a thickness $d_{\mathrm{Ge}} = 110$~nm. 
Germanium was chosen for several reasons: (i) its high refractive index ($n \approx 4.2$ at 1550 nm) enables strong optical confinement; (ii) its near-infrared absorption coefficient allows for efficient hot electron generation via internal photoemission from the metal; and (iii) its compatibility with silicon CMOS processing facilitates integration into established semiconductor technology. 
Figure~\ref{fig:fig_HMM_1}(b) presents the complex refractive index of Ge as a function of wavelength, as used in our simulations, with data reproduced from Nunley et al.~\cite{Nunley2016}.

The PhC consists of six ($N=6$) periods of alternating layers including a silicon (Si)~\cite{Green1995} layer and a HMM layer. 
The HMM is a sub-wavelength multilayer structure consisting of four periods of alternately deposited ITO and Si layers.
The entire structure is assumed to be fabricated on a BK7 optical glass substrate with refractive index n$_{\mathrm{sub}} = 1.5$ and illuminated from the top (air side) by a plane wave with a variable incidence angle $\theta$ and polarizations.

\subsection{Phase-Matching Condition}
\label{sec:phase_matching}

The excitation of a TPP at the metal-PhC interface can be explained by phase-matching condition. 
For a wave circulating in a cavity formed between a metal and a PhC, constructive interference requires the round-trip phase to accumulate an integer multiple of $2\pi$:

\begin{equation}
\varphi_{\mathrm{Total}}(\lambda,\theta) = \varphi_{\mathrm{M}}(\lambda,\theta) + 2\varphi_{\mathrm{top}}(\lambda,\theta) + \varphi_{\mathrm{PhC}}(\lambda,\theta) = 2m\pi,
\label{eq:phase_condition}
\end{equation}

\noindent where $m$ is an integer, typically $m=0$ for the fundamental TPP mode.

The three phase contributions have distinct physical origins:
\begin{itemize}
    \item $\varphi_\mathrm{M}$ is the phase shift upon reflection from the metal layer. In a good conductor at infrared wavelengths, $\varphi_\mathrm{M} \approx \pi$ with a weak angular dependence. Hence, we have $\partial\varphi_{\mathrm{M}}/\partial\theta\approx0$.
    \item $2\varphi_{\mathrm{top}}$ is the accumulated phase during propagation through the top Ge layer forwand and backward.
    \item $\varphi_{\mathrm{PhC}}$ is the phase of the reflection coefficient from the semi-infinite PhC, which depends on the band structure of the periodic medium.
\end{itemize}

\begin{figure}[t]
    \centering
    \includegraphics[width=80mm]{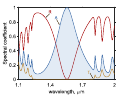}
    \caption{Reflectance (R), transmittance (T), and absorptance (A)  spectra of proposed structure at normal incidence of light.}
    \label{fig:Fig2}
\end{figure}
The phase accumulation in the top layer is explicitly given by:

\begin{equation}
2\varphi_{\mathrm{top}}(\lambda,\theta) = \frac{4\pi d_{\mathrm{top}}}{\lambda} \sqrt{\varepsilon_{\mathrm{top}} - \sin^2\theta},
\label{eq:top_phase}
\end{equation}

\noindent where $d_{\mathrm{top}} = d_{\mathrm{Ge}}$ and $\varepsilon_{\mathrm{top}} = \varepsilon_{\mathrm{Ge}}$. Differentiating with respect to $\theta$ at fixed $\lambda$ yields:

\begin{equation}
\frac{\partial \varphi_{\mathrm{top}}}{\partial \theta} = -\frac{2\pi d_{\mathrm{top}}}{\lambda} \frac{\sin (2\theta)}{\sqrt{\varepsilon_{\mathrm{top}} - \sin^2\theta}} < 0,
\end{equation}

\noindent confirming that $2\varphi_{\mathrm{top}}$ decreases monotonically with increasing $\theta$.

For a conventional all-dielectric PhC, the reflection phase $\varphi_{\mathrm{PhC}}$ also decreases with $\theta$. This can be understood by considering that the effective optical thickness of each layer decreases as $n_j d_j \cos\theta_j$, effectively shortening the cavity. 
Consequently, the total phase $\varphi_{\mathrm{Total}}$ decreases, requiring a shorter wavelength (blue shift) to restore the phase-matching condition ($\varphi_{\mathrm{Total}}=2m\pi$).

\subsection{Phase Compensation via HMM Integration}


When constituent layers in a multilayer structure are much thinner than the wavelength of light, the entire structure can be considered as a homogeneous effective medium with uniaxial anisotropy. This effective medium approximation (EMA) is valid when $d_{\mathrm{ITO}}n_\mathrm{ITO} + d_{\mathrm{Si}}n_\mathrm{Si}\ll\lambda$.

For a periodic stack of two materials with dielectric permittivities $\varepsilon_\mathrm{ITO}$ and $\varepsilon_\mathrm{Si}$ and volume fractions $f$ and $1-f$, respectively, the effective permittivity tensor components are given by:

\begin{equation}
\varepsilon_\mathrm{Bx} = f \varepsilon_{\mathrm{ITO}} + (1-f) \varepsilon_{\mathrm{Si}},
\label{eq:eps_parallel}
\end{equation}

\begin{equation}
\frac{1}{\varepsilon_\mathrm{Bz}} = \frac{f}{\varepsilon_{\mathrm{ITO}}} + \frac{1-f}{\varepsilon_{\mathrm{Si}}}.
\label{eq:eps_perpendicular}
\end{equation}

Here, $\varepsilon_\mathrm{Bx}$ is the in-plane (ordinary) component, and $\varepsilon_\mathrm{Bz}$ is the out-of-plane (extraordinary) component.

The dielectric permittivity of ITO is described by the Drude model:

\begin{equation}
\varepsilon_{\mathrm{ITO}}(\omega) = \varepsilon_{\infty} - \frac{\omega_p^2}{\omega^2 + i\gamma\omega},
\label{eq:drude}
\end{equation}

\noindent with parameters $\varepsilon_{\infty} = 3.9$, $\hbar\omega_p = 2.48$~eV, and $\hbar\gamma = 0.016$~eV, taken from Ref.~\cite{Gerfin1996}. 
These parameters yield a plasma wavelength $\lambda_p \approx 500$~nm and a crossover wavelength where  $\mathfrak{R}(\varepsilon_{\mathrm{ITO}}) = 0$ at approximately $\lambda \approx 1.2$~$\mu$m.

Figure~\ref{fig:fig_HMM_1}(c) shows the calculated real and imaginary parts of $\varepsilon_\mathrm{Bx}$ and $\varepsilon_\mathrm{Bz}$ as functions of wavelength. 
In the telecommunication wavelength range, $\mathfrak{R}(\varepsilon_\mathrm{Bx}) > 0$ while $\mathfrak{R}(\varepsilon_\mathrm{Bz}) < 0$, indicative of type-I hyperbolic dispersion. 
This regime is characterized by an open isofrequency surface, enabling high-density photonic states and anomalous phase accumulation.

For TM-polarized waves incident on a uniaxial medium with optical axis perpendicular to the interfaces, the effective refractive index for the extraordinary wave is \cite{Yariv1984}:

\begin{equation}
n_{\mathrm{eff}}(\theta) = \sqrt{\frac{\varepsilon_\mathrm{Bx}\varepsilon_\mathrm{Bz}}{\varepsilon_\mathrm{Bx}\sin^2\theta + \varepsilon_\mathrm{Bz}\cos^2\theta}}.
\end{equation}

In the type-I hyperbolic regime [$\varepsilon_\mathrm{Bx}>0$, $\varepsilon_\mathrm{Bz}<0$], the denominator can become small or even vanish at certain angles, leading to a rapid increase in $n_{\mathrm{eff}}$ with $\theta$. This, in turn, causes $\varphi_{\mathrm{PhC}}$ to increase with $\theta$, counteracting the decrease from $\varphi_{\mathrm{top}}$.

By carefully engineering the layer thicknesses, we can achieve a regime in which:

\begin{equation}
\frac{\partial \varphi_{\mathrm{PhC}}}{\partial \theta} \approx -\frac{\partial (2\varphi_{\mathrm{top}})}{\partial \theta},
\end{equation}

\noindent resulting in $\partial \varphi_{\mathrm{Total}}/\partial \theta \approx 0$ and an angle-insensitive resonance wavelength.

To realize a PhC with anomalous dispersion (a "red" shift of the band gap with increasing angle of incidence), the thicknesses of the HMM layers must satisfy the conditions proposed in~\cite{Wu2018}:

\begin{equation}
d_{\mathrm{Si}} = \frac{\lambda_{\mathrm{Brg}}/2 - d_{\mathrm{ITO}}\sqrt{\Re(\varepsilon_\mathrm{Bx})}}{\sqrt{\varepsilon_\mathrm{Si}}},
\label{eq:dA}
\end{equation}

\begin{equation}
d_{\text{HMM}} > d_{{\text{HMM}},\mathrm{min}} = \frac{\lambda_{\mathrm{Brg}}/2}{\sqrt{\Re(\varepsilon_\mathrm{Bx})}\left(1 - \frac{\varepsilon_\mathrm{Si}}{\Re(\varepsilon_\mathrm{Bz})}\right)},
\label{eq:dB}
\end{equation}

\noindent $\lambda_{\mathrm{Brg}}$ is the wavelength corresponding to the center of the PhC band gap.

For the proposed structure, \(d_{\text{Si}} = 131\)~nm and \(d_{\text{HMM}} = 168\)~nm. For a fixed filling factor \(f = 0.65\), the ITO layer thickness is \(d_{\text{ITO in HMM}} =fd_{\text{HMM}}/4= 27.3\)~nm, and the silicon layer within the HMM has a thickness \(d_{\text{Si in HMM}} = (1-f)d_{\text{HMM}}/4 = 14.7\)~nm.

\subsection{Critical Coupling Condition}

For a photodetector to perform optimally, it is essential not only that the resonance wavelength remains stable across angles of incidence, but also that the incident wave couples efficiently to the TPP mode. 
This condition is met when the structure satisfies critical coupling, i.e., when the radiative decay rate $\gamma_{\mathrm{rad}}$ equals the absorptive decay rate $\gamma_{\mathrm{abs}}$~\cite{Haus1983-ji,Konov2024}. 
Under critical coupling, the reflection vanishes and all incident light is absorbed by the structure.

In our design, we adjust the thickness of the Ti layer to achieve critical coupling at normal incidence. 
The thin metal film serves a dual purpose: it provides the necessary reflection phase shift to support the TPP mode and introduces the absorptive loss required for hot electron generation.
By keeping the Ti layer ultrathin ($d_{\mathrm{Ti}} = 7$~nm), we ensure that $\gamma_{\mathrm{abs}}$ remains moderate, allowing the mode to preserve a reasonably high quality factor while still enabling complete light absorption.

\subsection{Transfer Matrix Method}

All optical simulations in this work are performed using the transfer matrix method (TMM) for multilayer structures~\cite{Yeh1979}. 
This approach is particularly well-suited for one-dimensional layered geometries and provides exact solutions to Maxwell's equations for plane wave incidence.

For a multilayer stack with $N$ layers, the total transfer matrix $\mathbf{M}$ is obtained by multiplying the individual layer matrices:

\begin{equation}
\mathbf{M} = \mathbf{D}_0^{-1} \left( \prod_{j=1}^{N} \mathbf{D}_j \mathbf{P}_j \mathbf{D}_j^{-1} \right) \mathbf{D}_\mathrm{s},
\end{equation}

\noindent where $\mathbf{D}_j$ is the dynamical matrix for layer $j$, $\mathbf{P}_j$ is the propagation matrix accounting for phase accumulation across the layer thickness, and $\mathbf{D}_0$ and $\mathbf{D}_\mathrm{s}$ are the dynamical matrices for the incident and substrate media, respectively.

For a layer with refractive index $n_j$ and thickness $d_j$, the propagation matrix is diagonal:

\begin{equation}
\mathbf{P}_j = \begin{pmatrix}
e^{i k_{zj} d_j} & 0 \\
0 & e^{-i k_{zj} d_j}
\end{pmatrix},
\end{equation}

\noindent where $k_{zj} = (2\pi/\lambda) \sqrt{n_j^2 - \sin^2\theta}$ is the normal component of the wave vector in the layer.

The reflection and transmission coefficients of the entire structure are obtained from the elements of the total transfer matrix:

\begin{equation}
r = \frac{M_{21}}{M_{11}}, \quad t = \frac{1}{M_{11}}.
\end{equation}

The reflectance, transmittance, and absorptance are then given by $R = |r|^2$, $T = |t|^2$, and $A = 1 - R - T$, respectively.

For anisotropic layers such as the HMM, the TMM is extended to handle uniaxial media with optical axis perpendicular to the interfaces. In this case, the ordinary and extraordinary waves propagate independently, and the dynamical matrices are modified to account for the polarization-dependent refractive indices.

\begin{figure*}[t]
    \centering
    \includegraphics[width=\linewidth]{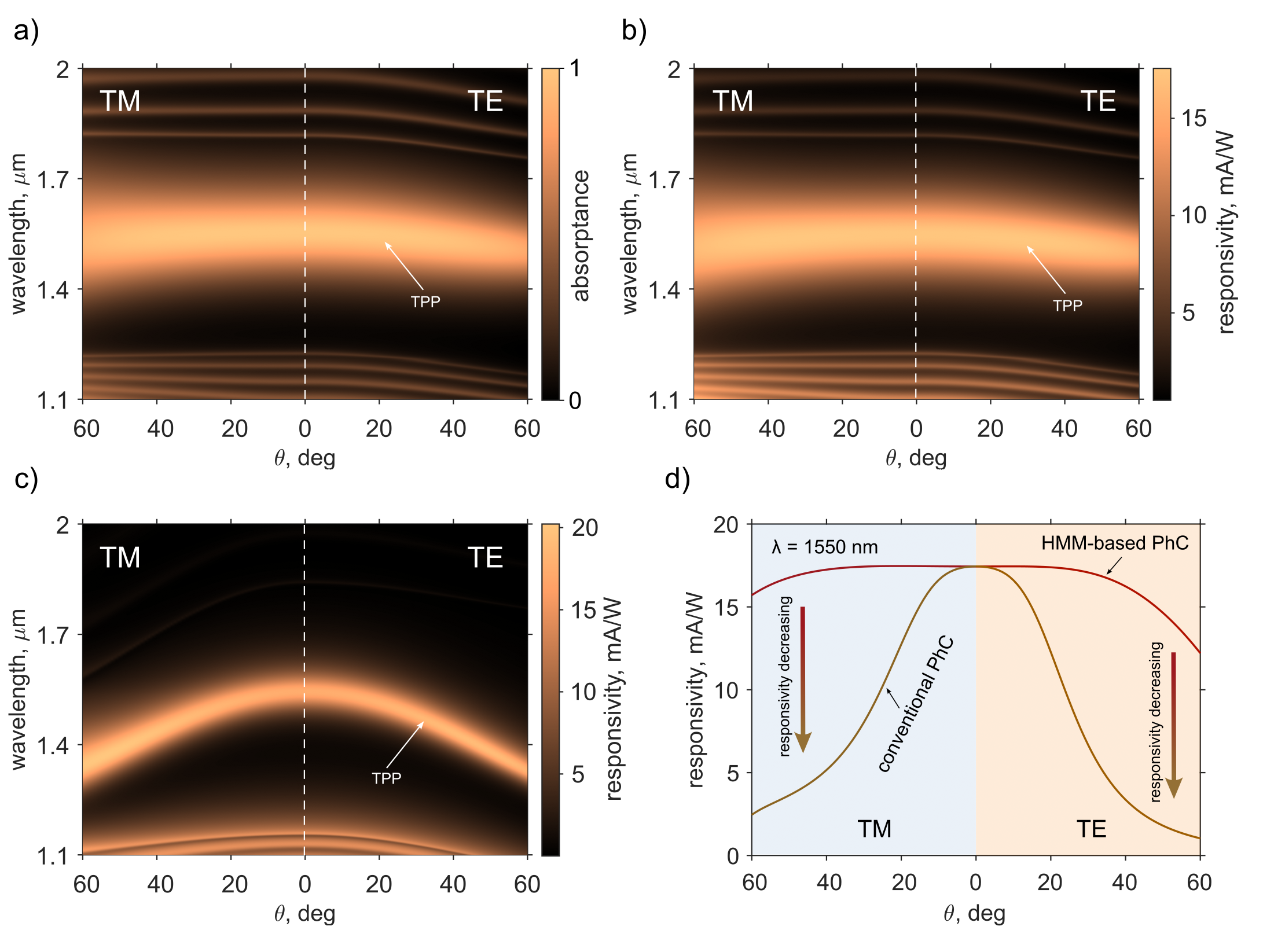}
    \caption{(a) Reflectance spectra of the structure as a function of wavelength and incidence angle for TM and TE polarizations.  (b) Responsivity of the structure as a function of wavelength and incidence angle for TM and TE polarizations.  (c) Responsivity of the structure without HMM as a function of wavelength and incidence angle for TM and TE polarizations. (d) The responsivities of the HMM-based structure and the structure without HMM are compared at a fixed wavelength and for different angles of incidence for TM and TE polarizations.}
    \label{Fig3}
\end{figure*}

\section{Results and Discussion}
\label{sec:results}

\subsection{Band Gap of the Bare PhC}

Figure~\ref{fig:fig_HMM_1}(d) shows reflectance maps for TM and TE polarizations calculated for the bare PhC without homogenization approximation. 
For TM polarization, the PhC band gap remains remarkably stable across the entire angular range from 0$^\circ$ to 60$^\circ$. 
This behavior is a direct consequence of the hyperbolic dispersion in the HMM layers, which compensates the angular dependence of the optical path length. 
For TE polarization, a slight blue shift of approximately 30~nm is observed at large angles. 
This is expected because TE-polarized waves experience only the ordinary refractive index of the HMM ($n_o = \sqrt{\varepsilon_\mathrm{Bx}}$), which does not exhibit the anomalous angular dependence of the extraordinary index. 
Nevertheless, the angular dispersion for TE polarization remains significantly smaller than in conventional all-dielectric PhCs.

\subsection{TPP Resonance}

Figure~\ref{fig:Fig2} shows the optical response of the complete structure including the Ti and Ge layers. 
At normal incidence, the reflectance spectrum exhibits a resonant dip at $\lambda = 1550$~nm, dropping to nearly zero. 
Correspondingly, the absorptance reaches 100\%, indicating that the critical coupling condition is satisfied. 
The transmittance is negligible throughout the band gap due to the high reflectivity of the PhC and the absorptive Ti layer. 

Figure~\ref{Fig3}(a) shows angular-dependent absorptance maps for TM and TE polarizations, respectively. 
For TM polarization, the TPP resonance remains centered at 1550~nm across the entire angular range from 0$^\circ$ to 60$^\circ$, with only a slight broadening at larger angles. 
The absorptance remains above 90\% for all angles up to 60$^\circ$, demonstrating the wide-angle operation of the device. 
For TE polarization, a small blue shift of approximately 15~nm is observed at $\theta = 60^\circ$, consistent with the band gap behavior shown in Fig.~\ref{fig:fig_HMM_1}(d). 
Nevertheless, the resonance remains well-defined and maintains high absorptance.

\subsection{Photodetector Responsivity}
The mechanism of incident radiation detection is described using the Fowler model of internal photoemission. 
Unlike the classical external photoelectric effect, the present scheme relies on internal photoemission of hot electrons across the Schottky potential barrier formed at the metal–semiconductor interface.
The efficiency of this process in plasmonic structures is governed by three interrelated factors: resonant optical absorption, hot electron transport within the metal, and the probability of overcoming the Schottky barrier.

High optical absorption at the TPP wavelength is essential for efficient conversion of incident radiation into electronic excitations. 
Photons resonantly absorbed in the ultrathin metal film transfer their energy to conduction electrons, generating nonequilibrium hot electrons. 
As these electrons migrate toward the interface with the semiconductor, they undergo scattering from defects, phonons, and other carriers. 
Selecting a metal film with a thickness below the electron mean free path minimizes transport losses, ensuring that a substantial fraction of electrons reach the barrier with sufficient energy to overcome it.

Electrons arriving at the metal–semiconductor interface encounter the Schottky potential barrier. 
Only those with kinetic energy exceeding the barrier height can be injected into the conduction band of germanium. 
The probability of this process is described by the Fowler model of internal photoemission. 
Once injected into the contact layer of indirect-bandgap germanium—where the absorption coefficient increases quadratically from the band edge—the electrons are collected by the external circuit, generating a photocurrent $I_\mathrm{ph}$.

By combining the sequential stages of energy conversion, an expression for the photosensitivity $R(\lambda,\theta)$ can be rigorously derived. 
The total photocurrent is given by the product of the incident photon flux $N_\mathrm{ph} = P_{\text{opt}}/(h\nu)$, the resonant absorption $A(\lambda,\theta)$, the elementary charge $q$, and the probabilities of transport to the barrier and subsequent injection. By introducing the total internal quantum efficiency $\eta(\lambda)$, which encompasses both the transport and injection stages,  the photosensitivity can be expressed according to the Fowler model \cite{PhysRev.38.45}:

\begin{equation}
    R = \frac{I_\mathrm{ph}}{P_{\text{opt}}} = \frac{q A(\lambda,\theta) \eta(\lambda)}{h \nu}, \quad \eta(\lambda) = \frac{(h\nu - \Delta E_\mathrm{b})^2}{4 E_\mathrm{f} h \nu},
\label{eq.fowler}
\end{equation}

\noindent where $\Delta E_\mathrm{b}$ is the Schottky barrier height at the Ti/Ge interface and $E_\mathrm{f}$ is the Fermi energy of titanium.
For the Ti/Ge interface, we use $\Delta E_\mathrm{b} = 0.36$~eV and $E_\mathrm{f} =~ 4.33$~eV, based on literature values~\cite{C9TC04345D, rumble2023crc}.

\begin{table*}[t]
\centering
\caption{Comparison of HMM-based and conventional TPP photodetectors.}
\label{tab:comparison}
\begin{tabular}{l c c}
\hline
\textbf{Parameter} & \textbf{HMM-based PhC} & \textbf{Conventional PhC} \\
\hline
Normal-incidence responsivity (mA/W) & 17.5 & 17.5\\
Resonance wavelength (nm) & 1550 & 1550 \\
TM angular tolerance (60$^\circ$ drop) & 10\% & 86\% \\
TE angular tolerance (60$^\circ$ drop) & 30\% & 94\% \\
Angular range for $>90\%$ of peak (TM) & 0$^\circ$--55$^\circ$ & 0$^\circ$--15$^\circ$ \\
Number of periods $N$ & 6 & 10 \\
\hline
\end{tabular}
\end{table*}

Figure~\ref{Fig3}(b) presents the calculated responsivity spectra for TM polarization at various incidence angles. 
At normal incidence, the responsivity reaches 17.5~mA/W at $\lambda = 1550$~nm. 
As the angle increases to 60$^{\circ}$, the responsivity slightly decreases to 15.8 mA/W, with a reduction of about 10\%.
Notably, the spectral position of the peak remains fixed at 1550~nm, confirming the angle-insensitive design. 
For TE polarization, the responsivity decreases more rapidly, dropping to 12.2~mA/W at 60$^\circ$ (a 30\% reduction), consistent with the slight blue shift observed in the absorptance.

\subsection{Comparison with Conventional Design}
\label{sec:comparison}

To highlight the advantages of our HMM-based approach, we compare its performance with that of a conventional TPP photodetector based on an all-dielectric PhC. 
The reference structure consists of a PhC with alternating SiO$_2$ \cite{Malitson:65} and Si layers, each 140~nm thick, and $N = 10$ periods. 
The top SiO$_2$ spacer thickness is $d_{\mathrm{top}} = 375$~nm, while the Ge (110~nm) and Ti (7~nm) layers are identical to those in the proposed structure. 
This reference design is optimized to achieve near-unity absorption at 1550~nm under normal incidence, as shown in Fig.~\ref{Fig3}(c).

Figure~\ref{Fig3}(d) compares the angular stability of the two designs. 
For the conventional PhC, the responsivity drops sharply with increasing angle. 
At $\theta = 60^\circ$, the TM-polarized responsivity is reduced by 86\% relative to its normal-incidence value, while the TE-polarized responsivity is reduced by 94\%. 
This severe degradation, typical of TPP-based devices, arises from the resonance shifting out of the fixed detection band. 
In contrast, the HMM-based PhC retains 90\% of its TM responsivity and 70\% of its TE responsivity at $60^\circ$. 
The improvement is particularly pronounced for TM polarization, for which the design is optimized to fully exploit the hyperbolic dispersion.

Table~\ref{tab:comparison} summarizes the key performance metrics for both structures.

\section{Conclusion}
\label{sec:conclusion}

In summary, we have theoretically demonstrated a TPP-based photodetector with exceptional angular tolerance achieved through hyperbolic metamaterial engineering. 
By incorporating ITO/Si multilayer HMMs into a one-dimensional PhC, we compensated the natural blue shift of the TPP resonance, pinning it at the telecommunication wavelength of 1550~nm over a wide range of incidence angles. 
The HMM-based PhC exhibits an angle-independent band gap for TM polarization, with the center wavelength varying by less than 10~nm from 0$^\circ$ to 60$^\circ$. 
The calculated photodetector responsivity at normal incidence is 17.5~mA/W, decreasing by only 10\% at $\theta = 60^\circ$ for TM polarization. 
This represents a dramatic improvement over conventional all-dielectric PhC designs, which exhibit 86\% degradation under the same conditions.

Our results establish HMM-engineered TPPs as a robust platform for wide-angle high-performance optoelectronic devices and open new avenues for dispersion engineering in active plasmonic systems. 
The design principles demonstrated here can be extended to other spectral ranges and material platforms, with potential applications in imaging, sensing, energy harvesting, and quantum photonics.

\acknowledgments
The work was carried out within the state assignment
of the Federal Research Center KSC SB RAS.

\bibliographystyle{apsrev4-2}
\bibliography{library}

\end{document}